\documentclass[sigconf]{acmart}

\usepackage{subcaption}
\usepackage{listings}
\usepackage{xcolor}
\usepackage{colortbl}
\usepackage{algorithm}
\usepackage{algorithmic}
\usepackage{multirow}
\usepackage{enumitem}
\usepackage{svg}
\setlist[itemize]{leftmargin=1em}

\lstdefinestyle{sdc}{
    language=tcl,
    basicstyle=\ttfamily\footnotesize,
    keywordstyle=\color{blue},
    commentstyle=\color{gray},
    stringstyle=\color{orange},
    breaklines=true,
    frame=single,
    numbers=left,
    numberstyle=\tiny\color{gray},
    captionpos=b
}
\AtBeginDocument{%
  }

\copyrightyear{2026}
\acmYear{2026}
\setcopyright{cc}
\setcctype{by}
\acmConference[GLSVLSI '26]{Great Lakes Symposium on VLSI 2026}{June 22--24, 2026}{Canandaigua, NY, USA}
\acmBooktitle{Great Lakes Symposium on VLSI 2026 (GLSVLSI '26), June 22--24, 2026, Canandaigua, NY, USA}
\acmDOI{10.1145/3787109.3815265}
\acmISBN{979-8-4007-2431-2/2026/06}
\acmConference[GLSVLSI '26]{Great Lakes Symposium on VLSI}{June 22--24,
  2026}{Finger Lakes, NY}

\begin{document}

\title{An Open-Source Flow for Single-Phase, Edge-Triggered to Two-Phase, Non-Overlapping Clocking Conversion}

\author{Paolo Pedroso}
\affiliation{%
	\institution{University of California Santa Cruz}
	\city{Santa Cruz}
	\state{California}
	\country{USA}}
\email{ppedroso@ucsc.edu}

\author{Lee-Way Wang}
\affiliation{%
	\institution{University of California Santa Cruz}
	\city{Santa Cruz}
	\state{California}
	\country{USA}}
\email{lwang261@ucsc.edu}

\author{Matthew R. Guthaus}
\affiliation{%
	\institution{University of California Santa Cruz}
	\city{Santa Cruz}
	\state{California}
	\country{USA}}
\email{mrg@ucsc.edu}


\begin{abstract}
	Two-phase clocking offers significant advantages in timing margin and clock flexibility, yet its adoption remains limited due to the absence of automation in modern design flows. Managing strict non-overlap and 180$^\circ$ phase separation introduces complexity in RTL implementation and timing closure, leaving two-phase clocking rare in practice. This paper presents the first fully automated two-phase clocking flow integrated into OpenROAD Flow Scripts (ORFS). Our methodology automatically transforms flip-flop-based RTL into two-phase latch-based designs using Yosys technology mapping, ABC retiming, dual clock tree synthesis, two-phase correctness validation, and full physical design from RTL-to-GDS. We implement clock-gated and recirculation mux variants, where clock-gated achieves an average 29.2\% power reduction and 50\% latch count reduction over recirculation mux. Both variants are compared against flip-flop baselines, demonstrating timing closure through time borrowing on a design that failed timing with flip-flops.
\end{abstract}

%




\maketitle

\section{Introduction}
\label{sec:intro}

The two-phase clocking scheme dates back to the early 1970s, with early microprocessors such as the Intel 8008 among its first prominent implementations. As technology nodes shrank and operating frequencies increased through the 1980s, two-phase systems introduced complexities in timing verification, particularly concerning hold time violations, race conditions, and precise control of non-overlapping clock phases~\cite{Hurst2006}.  Managing clock skew with multiple phases then became increasingly difficult as circuit complexity rose~\cite{Friedman2001}. By the time logic synthesis, automated physical design, and modern STA became standard practice, two-phase clocking was no longer a focus, and the entire EDA ecosystem shifted to single-phase, edge-triggered flip-flop designs. Two-phase clocking is rarely found in modern integrated circuits, and the industry's widespread adoption of flip-flop-based designs combined with EDA tools optimized for this paradigm has made two-phase systems largely obsolete. The dominance of single-phase, edge-triggered flip-flops has left two-phase clocked latches largely overlooked in modern digital design, a gap that warrants renewed investigation. Two-phase clocking offers additional benefits including flexibility, reduced clock skew sensitivity, and time borrowing capabilities that enable higher operating frequencies~\cite{Horowitz271Clocking}. The paradigm allows tunability, especially for fixing hold times, which can cause permanent failure. For example, many of the open-source designs on the Google-Skywater shuttle run failed because of hold time problems in the GPIO scan chain. In two-phase designs, widening the $\Phi_1$ and $\Phi_2$ gap relaxes the hold constraint without the cost of reducing clock frequency.


Latches themselves offer inherent power and area advantages over flip-flops~\cite{Hurst2006}, as they use simpler level-sensitive circuits with fewer transistors. The widespread dominance of flip-flop-based designs, combined with the availability of open-source EDA tools such as OpenROAD, Yosys, and ABC~\cite{OpenROAD2026, ajayi2019openroad, Yosys2013, Brayton2010}, presents a unique opportunity to democratize access to unconventional design methodologies. By leveraging these tools, we introduce two-phase clocking into the OpenROAD ecosystem, enabling the designer to transform flip-flop-based RTL netlists into two-phase non-overlapping latch-based implementations without manual expert intervention, making an otherwise inaccessible design paradigm openly available for research, education, and exploration.
Our contributions include:
\begin{itemize}
	\item \textbf{Latch Equivalents:} We developed a set of sky130hd latch-based equivalents of standard flip-flop variants, including synchronous and asynchronous resets, sets, and enables, implemented as functionally verified two-phase latches in both clock-gated and
	      recirculation mux variants.

	\item \textbf{Synthesis Methodology:} We propose a flip-flop to two-phase latch synthesis methodology based on smart retiming with ABC and Yosys technology mapping, with equivalence checking via Yosys~\cite{Singh2018, Hurst2007, YosysHQ2024Techmap, YosysEquiv} before and after retiming.

	\item \textbf{Dual Clock Tree Synthesis:} We perform dual clock tree synthesis using TritonCTS~\cite{TritonCTS} for the $180^\circ$ out-of-phase, non-overlapping $\Phi_1$ and $\Phi_2$ clocks, with specific options to account for differing latch insertion delays, fix long wires before latency adjustment, and prevent clock buffers from being
	      placed on blockages.

	\item \textbf{Two-Coloring Static Verification:} We developed a two-coloring static verification pass~\cite{Wolf2009VLSI} to validate two-phase latch implementations at the finishing step.
\end{itemize}

Together, these contributions provide the first open, automated path from a flip-flop-based RTL design to a fully implemented two-phase clocked system, making two-phase clocking accessible and reproducible for the broader research community. The remainder of this paper is organized as follows. Section~\ref{prelim} goes over the timing constraints of two-phase clocking, flip-flop and latch variant equivalents and retiming. Section~\ref{sec:frontend} proposes the frontend of the methodology while Section~\ref{sec:backend} goes over the backend. Section~\ref{sec:expsetup} and Section~\ref{expresults} present the experimental setup and results respectively.


\section{Preliminary}
\label{prelim}


\subsection{Timing Constraints}

In flip-flop designs, setup violations can be resolved by reducing the clock frequency, but hold violations cannot since hold time is independent of the clock period. Two-phase clocking offers more flexibility, because widening the spacing between the two clock pulses improves hold slack at the cost of some setup slack. This tunability makes two-phase clocking more resilient to clock skew and hold time issues than flip-flop-based designs. Another advantage of level-sensitive latches over flip-flops is time borrowing, in which a stage that cannot meet timing borrows slack from the next stage. Harris et al.~\cite{Harris1999} formalize the timing constraints of latch-based systems, presenting the setup requirement:
\begin{equation}
	A^c_i + \Delta_{DCi} + t^{pi,c}_{skew} \leq T_{P_i}
	\label{equation:setup}
\end{equation}
where $A^c_i$ is the arrival time of a signal launched by clock $c$ at the input to latch $i$, $\Delta_{DCi}$ is the setup time requirement for latch $i$ between the data pin and the falling clock edge, $t^{pi,c}_{skew}$ is the skew between clock $c$ and the clock phase $p_i$ of latch $i$, and $T_{P_i}$ is the time at which the sampling clock edge arrives. As Eq.~\eqref{equation:setup} shows, slowing the clock causes the sampling edge to arrive later, relaxing the setup constraint. Time borrowing from the next latch has the same effect, as it also delays the sampling clock edge.

As for the hold requirement:
\begin{equation}
	\delta_{DQi} + \delta_{ji} + S_{p_jp_i} \geq T_i + \Delta_{CDi} + t^{pi,pj}_{skew} - T_c
	\label{equation:hold}
\end{equation}
where $\delta_{DQi}$ is the minimum input-to-output propagation delay through latch $i$, $\delta_{ji}$ is the minimum combinational delay between latches $j$ and $i$, $S_{p_jp_i}$ is the phase shift between $p_j$ and $p_i$, $T_i$ is the clock pulse width, $\Delta_{CDi}$ is the hold time requirement, $t^{pi,pj}_{skew}$ is the skew between phases $p_i$ and $p_j$, and $T_c$ is the clock period~\cite{Harris1999}. As Eq.~\eqref{equation:hold} shows, decreasing $T_i$ (i.e., widening the spacing between clock phases) relaxes the hold constraint, while time borrowing relaxes the setup constraint~\cite{Harris1999}.



\begin{figure}[h]
	\centering
	\includegraphics[width=\linewidth]{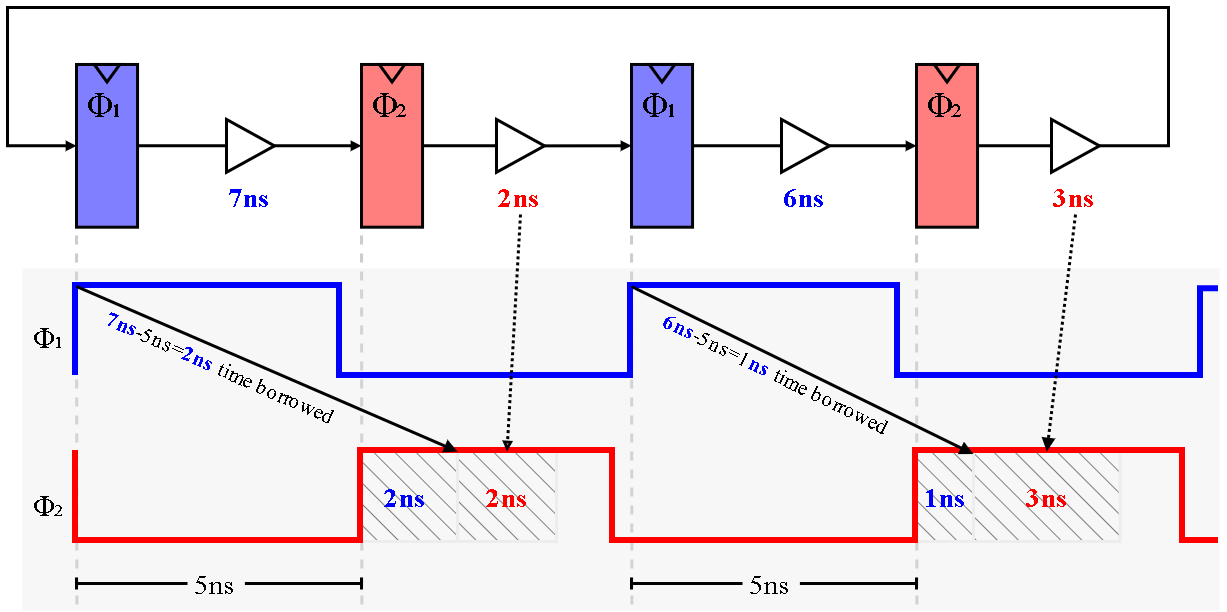}
	\caption{Time-borrowing example.}
	\Description{Time-borrowing example with $\Phi_1$ and $\Phi_2$ periods of 10ns.}
	\label{fig:time-borrowing}
\end{figure}
\subsection{Time Borrowing}

Time-borrowing is a technique enabled by two-phase clocking that can only be applied to latches. Because latches are level-sensitive, if a combinational path cannot meet timing to the next stage, the current stage can \textit{borrow} time from the next stage (assuming non-overlapping clocks). As shown in Fig.~\ref{fig:time-borrowing}, with a 10ns period for both $\Phi_1$ and $\Phi_2$, if the first stage has a combinational path of 7ns, it can borrow 2ns from the next stage; since the next stage's combinational path is only 2ns, timing is still met without race conditions.


\subsection{Retiming}
\label{retime}

Since our framework involves a series of technology mapping and retiming steps that progressively transform and rename cells throughout the flow, we introduce the following naming conventions similar to Yosys~\cite{YosysSimCells} to track cell types across each stage:
\begin{itemize}
	\item \textit{DFF\_base}$_{\Phi_n}$: Base posedge D flip-flop(s) (i.e.\ \textit{\_DFF\_P\_}) prior to mapping to latches, where $n$ denotes the clock phase ($\Phi_1$ or $\Phi_2$).
	\item \textit{DFF\_fvar}$_{\Phi_n}$: Full-variant posedge D flip-flop(s) (e.g.\ \textit{\_DFFE\_PP\_}) prior to mapping to \textit{DFF\_base}$_{\Phi_n}$, where $n$ denotes the clock phase ($\Phi_1$ or $\Phi_2$).
\end{itemize}

Retiming moves registers across combinational logic nodes while preserving output functionality. The algorithm guarantees that every path through the combinational network passes through the register boundary exactly once~\cite{Hurst2006}.

Rather than retiming directly on latch-based designs, it is more practical to first duplicate flip-flops, retime the duplicated flip-flops, and then later transform them into latch-based implementations, inspired by Singh et al.~\cite{Singh2018}. This is safe because a \textit{\_DFF\_P\_} is functionally equivalent to a two-phase latch pair, if both latches are positive-enable. The \textit{DFF\_base}$_{\Phi_1}$ captures the state, and the \textit{DFF\_base}$_{\Phi_2}$ outputs the state. Since the two clocks are non-overlapping and $180^\circ$ out of phase, the \textit{DFF\_base}$_{\Phi_2}$ is dependent on the output state of the \textit{DFF\_base}$_{\Phi_1}$, allowing the \textit{DFF\_base}$_{\Phi_1}$ to be freely retimed across combinational logic behind \textit{DFF\_base}$_{\Phi_2}$ without violating the phase relationship.

\begin{figure}[h]
	\centering
	\begin{subfigure}[b]{0.49\linewidth}
		\centering
		\includegraphics[width=\linewidth]{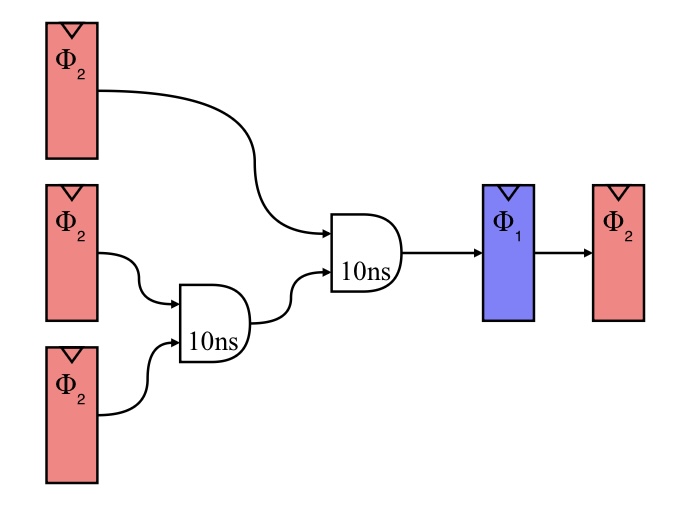}
		\caption{}
		\Description{Circuit diagram showing register placement before minimum-delay retiming.}
		\label{fig:before-mindelay}
	\end{subfigure}
	\hfill
	\begin{subfigure}[b]{0.49\linewidth}
		\centering
		\includegraphics[width=\linewidth]{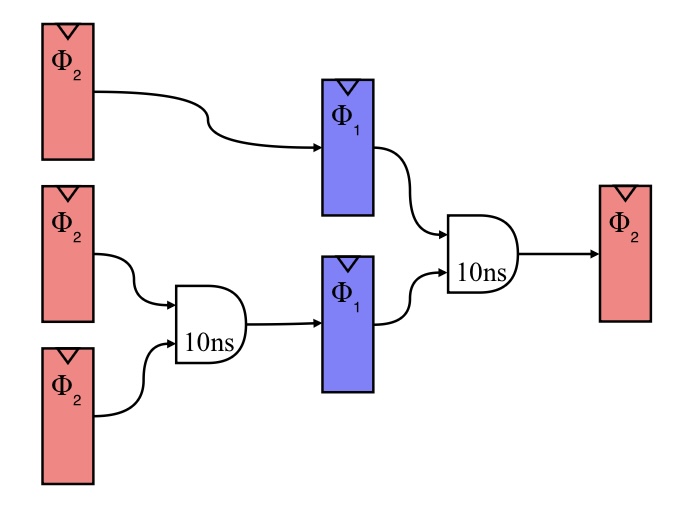}
		\caption{}
		\Description{Circuit diagram showing register placement after minimum-delay retiming.}
		\label{fig:after-mindelay}
	\end{subfigure}
	\caption{Minimum-delay retiming: (a) before and (b) after redistributing registers across combinational logic.}
	\label{fig:mindelay}
\end{figure}

\textbf{Minimum-delay retiming} Minimum-delay retiming redistributes registers forward across combinational logic to balance path depths across phases, reducing the critical path and contributing to frequency improvement. As shown in Fig.~\ref{fig:before-mindelay}, the critical path is 20ns prior to retiming, yielding a maximum frequency of 50MHz. After minimum-delay retiming, Fig.~\ref{fig:after-mindelay} demonstrates that the critical path is halved, achieving a maximum frequency of 100MHz.

\begin{figure}[h]
	\centering
	\begin{subfigure}[b]{0.49\linewidth}
		\centering
		\includegraphics[width=\linewidth]{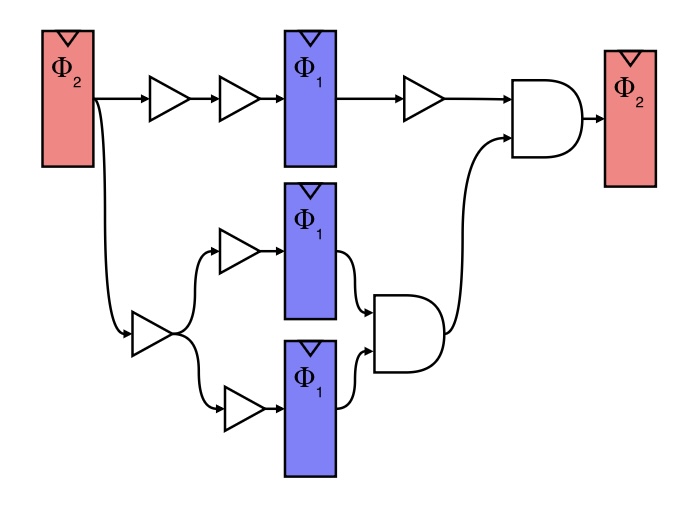}
		\caption{}
		\Description{Circuit diagram showing register placement before minimum-area retiming.}
		\label{fig:before-minarea}
	\end{subfigure}
	\hfill
	\begin{subfigure}[b]{0.49\linewidth}
		\centering
		\includegraphics[width=\linewidth]{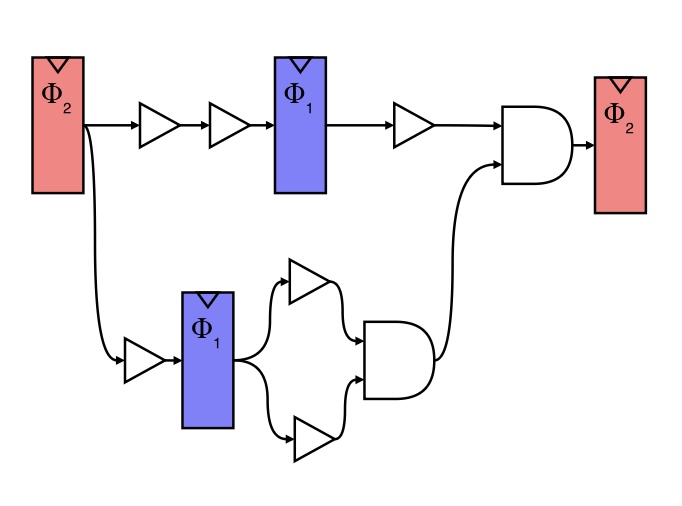}
		\caption{}
		\Description{Circuit diagram showing register placement after minimum-area retiming.}
		\label{fig:after-minarea}
	\end{subfigure}
	\caption{Minimum-area retiming: (a) before and (b) after minimizing total flip-flop count.}
	\label{fig:minarea}
\end{figure}

\textbf{Minimum-area retiming} Minimum-area retiming consolidates the register boundary to minimize total flip-flop count while maintaining the legal $\Phi_1$/$\Phi_2$ phase separation at the new register positions. As observed in Fig.~\ref{fig:before-minarea}, registers are distributed across multiple stages of the combinational logic path. After minimum-area retiming, Fig.~\ref{fig:after-minarea} demonstrates that the register count is reduced by merging boundaries, while the $\Phi_1$/$\Phi_2$ phase separation is preserved at the consolidated register positions.


\section{Methodology}
\begin{figure}[h]
	\centering
	\includegraphics[width=\linewidth]{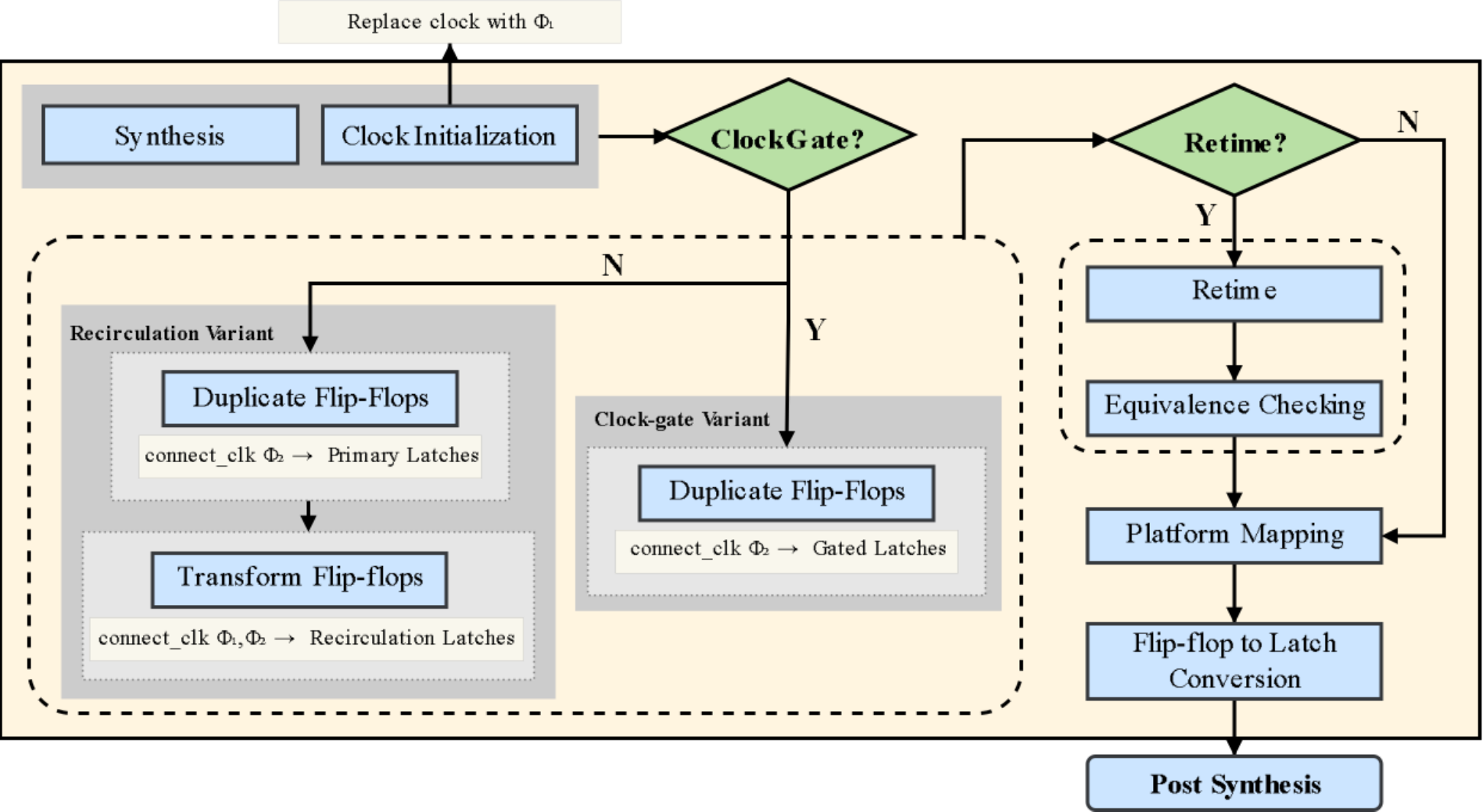}
	\caption{Two-phase front-end flow.}
	\Description{Diagram of two-phase front-end flow.}
	\label{fig:twophase-flow}
\end{figure}
\subsection{Frontend: Synthesis and Clock Conversion}
\label{sec:frontend}
The front-end uses Yosys \textit{techmap} to duplicate and transform flip-flops, and ABC \textit{retime} to redistribute sequential elements after duplication ~\cite{YosysHQ2024Techmap, Hurst2007}. We present the full frontend methodology in Fig.~\ref{fig:twophase-flow}. We use the \textit{\_DFFE\_P\_} cell, which is a positive edge D-type flip-flop with positive polarity enable, as an example throughout this section to illustrate the transformation to its two-phase clocked counterpart.

We also introduce a \textit{connect\_clk} command that manipulates the RTL netlist by reassigning module ports. The command operates on the post-synthesis netlist by enumerating all instances of a given cell type via Yosys \textit{select} and rewiring each instance's clock pin to the specified top-level clock port using Yosys \textit{connect}; it is called separately for $\Phi_1$ and $\Phi_2$ cell groups, ensuring that \textit{DFF\_base}$_{\Phi_1}$ instances are driven by \textit{clk\_1} and \textit{DFF\_base}$_{\Phi_2}$ instances by \textit{clk\_2} without modifying any combinational logic.

\textbf{Clock Port Initialization} To initialize a single-phase system into two-phase system we must add a second clock to the top-level. Prior to technology mapping, the design undergoes standard Yosys synthesis with two modifications: the existing clock port is renamed to \textit{clk\_1} ($\Phi_1$), and a second clock input \textit{clk\_2} ($\Phi_2$) is added to the top-level module.

\begin{figure*}[h]
	\centering
	\begin{subfigure}[b]{0.32\linewidth}
		\centering
		\includegraphics[width=\linewidth, height=4cm, keepaspectratio]{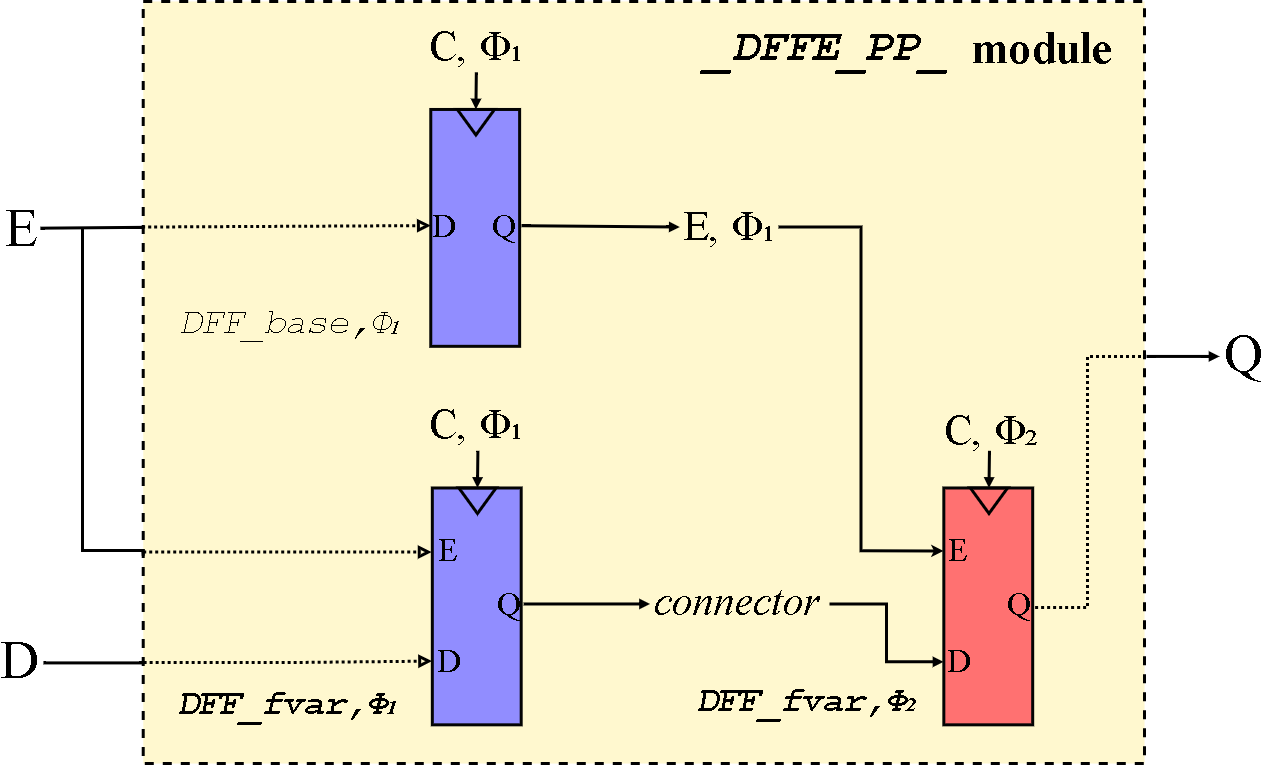}
		\caption{}
		\Description{Circuit diagram showing the recirculation mux duplication step.}
		\label{fig:recirc-dupe}
	\end{subfigure}
	\hfill
	\begin{subfigure}[b]{0.32\linewidth}
		\centering
		\includegraphics[width=\linewidth, height=4cm, keepaspectratio]{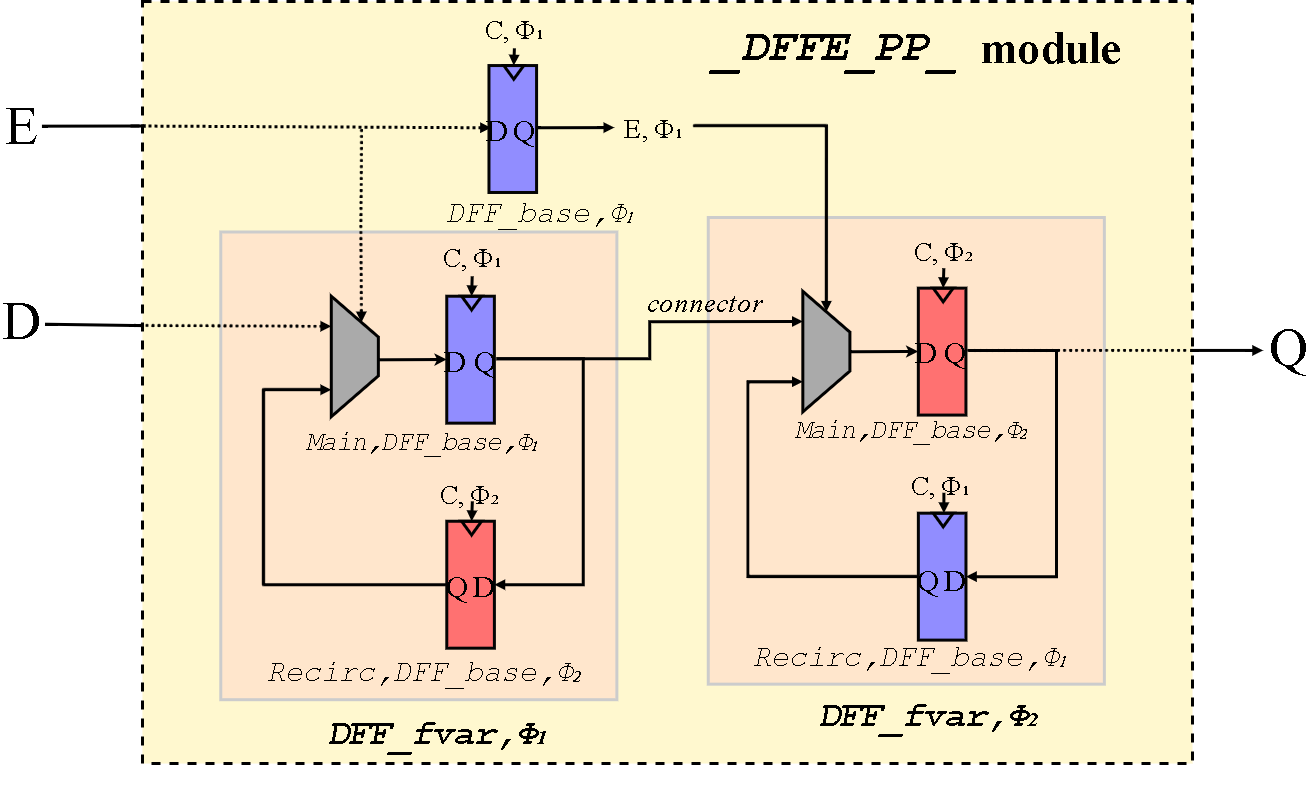}
		\caption{}
		\Description{Circuit diagram showing the recirculation mux final transformation.}
		\label{fig:recirc-transform}
	\end{subfigure}
	\hfill
	\begin{subfigure}[b]{0.32\linewidth}
		\centering
		\includegraphics[width=\linewidth, height=4cm, keepaspectratio]{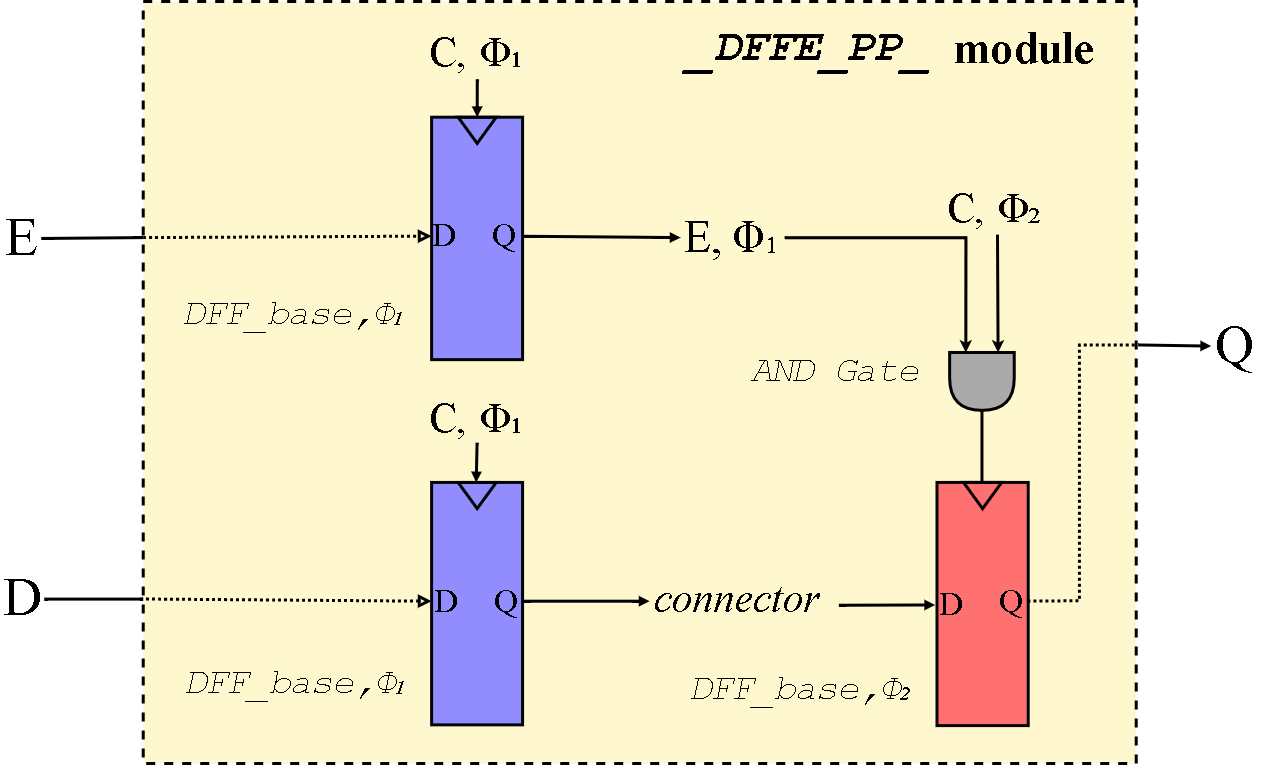}
		\caption{}
		\Description{Circuit diagram showing the clock-gated two-phase duplication/transformation.}
		\label{fig:cg-transform}
	\end{subfigure}
	\caption{Recirculation mux duplication (a), recirculation mux transformation (b), clock-gated duplication/transformation (c).}
	\label{fig:recirc-cg}
\end{figure*}


\textbf{Recirculation Mux Design: Duplicate D Flip-Flops} We perform a Yosys \textit{techmap} across all flip-flops in the design by duplicating each flip-flop to establish the foundation for retiming and for connecting clocks~\cite{YosysHQ2024Techmap}. Duplication means both flops in the duplicated pair retain their full variant type (i.e. \textit{DFF\_fvar}$_{\Phi_1,\Phi_2}$) as illustrated in Fig.~\ref{fig:recirc-dupe}. An additional \textit{DFF\_base}$_{\Phi_1}$ is inserted for the enable (\textit{E}) connecting it to the data input of \textit{DFF\_base}$_{\Phi_1}$, thus making \textit{E}$\rightarrow$\textit{E,$\Phi_1$} which connects to the second \textit{DFF\_fvar}$_{\Phi_2}$ control pin(s) to ensure that enable, reset, and set are valid signals fanning out to the correct phase, preventing control signals from propagating between different phases. By construction, \textit{E,$\Phi_2$} connects to \textit{DFF\_fvar}$_{\Phi_2}$, ensuring control signals do not cross phase boundaries.

\textbf{Recirculation Mux Design: Transform D Flip-Flops} Since the duplicated flops retain their full-variant type, they require further transformation to achieve their two-phase latch equivalents, as shown in Fig.~\ref{fig:recirc-transform}. Both \textit{DFF\_fvar}$_{\Phi_1,\Phi_2}$ are transformed into recirculation mux latches. Each \textit{DFF\_fvar}$_{\Phi_1,\Phi_2}$ is replaced by a \textit{Main,DFF\_base}$_{\Phi_1,\Phi_2}$ and a \textit{Recirc,DFF\_base}$_{\Phi_1,\Phi_2}$ pair connected via a 2:1 multiplexer. Input data \textit{D} feeds into the multiplexer, which drives \textit{Main,DFF\_base}$_{\Phi_1,\Phi_2}$, while \textit{Recirc,DFF\_base}$_{\Phi_1,\Phi_2}$ feeds back to hold the current state. The control pin of the multiplexer determines whether the output updates with new data or recirculates the existing state.



\textbf{Clock-Gated Design: Duplicate and Transform D Flip-Flops} Rather than preserving the full flip-flop variant type, all flip-flop variants are duplicated with an equivalent \textit{DFF\_base}$_{\Phi_1,\Phi_2}$. Functional equivalence with the original full-variant flip-flop is achieved by gating the clock input of \textit{DFF\_base}$_{\Phi_2}$, as shown in Fig.~\ref{fig:cg-transform}. An additional \textit{DFF\_base}$_{\Phi_1}$ is required to drive the control pins of \textit{DFF\_base}$_{\Phi_2}$ through an \textit{AND} gate, ensuring control signals are correctly phase-aligned.

\textbf{Retime} As described in Section~\ref{retime}, ABC \textit{retime} is applied to the duplicated flip-flop pairs prior to latch transformation. For the recirculation mux design, \textit{retime} is applied to \textit{DFF\_fvar}$_{\Phi_1,\Phi_2}$, redistributing the full-variant flip-flop pairs across combinational logic while maintaining the legal $\Phi_1$/$\Phi_2$ phase separation. For the clock-gated design, \textit{retime} is applied to \textit{DFF\_base}$_{\Phi_1,\Phi_2}$, redistributing the base flip-flop pairs across combinational logic.

\textbf{Equivalence Checking} Yosys equivalence checking~\cite{YosysEquiv} is performed on the design before and after retiming to confirm that retiming does not logically alter the circuit.

\textbf{DFF to Latch} In either variant, a final \textit{techmap} converts all \textit{DFF\_base}$_{\Phi_1,\Phi_2}$ into the platform's positive-enable latch.

\subsection{Backend: P\&R, CTS and Verification}
\label{sec:backend}
We connect the $\Phi_1$/$\Phi_2$ clocks during clock tree synthesis (CTS) with TritonCTS~\cite{TritonCTS} in OpenROAD~\cite{OpenROAD2026} with additional options to reduce clock skew, followed by two-phase correctness validation with OpenROAD's Python API~\cite{OpenROADAPI} to ensure \textit{all} latches are connected to opposite phases via a two-coloring check~\cite{Wolf2009VLSI}.

\textbf{Clock Tree Synthesis.} We build independent $\Phi_1$/$\Phi_2$ clock trees using \textit{clk\_nets} with TritonCTS. We use the options \textit{balance\_levels}, \textit{repair\_clock\_nets}, and \textit{obstruction\_aware} together to amend for the different insertion delays for $\Phi_1$/$\Phi_2$ latches. \textit{balance\_levels} attempts to keep a similar number of levels in the clock tree across non-register cells (e.g., clock-gate or inverter). \textit{repair\_clock\_nets} fixes long wires inside CTS prior to latency adjustment with delay buffers, which can lead to a more balanced clock tree. \textit{obstruction\_aware} enables obstruction-aware buffering such that clock buffers are not placed on top of blockages or hard macros.

\textbf{Two-coloring Static Verification.}
Two-coloring static verification checks whether a design satisfies two-phase discipline by assigning one color to all $\Phi_1$-derived signals and a second to all $\Phi_2$-derived signals~\cite{Wolf2009VLSI}. The coloring propagates through combinational logic, but alternates across latch boundaries, since each latch transfers data from one phase domain to the other.
A valid two-phase system must satisfy:
\begin{itemize}
	\item No latch may have an input and output signal of same color.
	\item The latch input signal and clock signal must be of same color.
	\item All signals to combinational logic element must be of same color.
\end{itemize}
The first rule ensures that a latch always transfers data between opposite phases, preventing same-phase transparency that could lead to races. The second rule guarantees that the data presented at the latch input is aligned with the clock phase that controls its transparency window. The third rule enforces phase separation in combinational logic so that signals from different clock domains are not mixed within the same logic cone.

\begin{algorithm}
	\caption{Build Latch Graph}
	\label{alg:buildgraph}
	\begin{algorithmic}[1]
		\STATE \textbf{Function} \textsc{LatchGraph}(\textit{design})
		\FORALL{instance $u$ in design}
		\IF{$u$ is sequential}
		\STATE Determine clock domain of $u$ via backwards DFS on clock net
		\STATE Add $u$ as node to $G$ with color($u$) $\gets$ $\Phi_1$ or $\Phi_2$
		\ENDIF
		\ENDFOR
		\FORALL{latch node $u$ in $G$}
		\STATE $visited\_nets \gets \emptyset$
		\STATE $F \gets$ DFS through combinational logic from $u$, marking traversed nets in $visited\_nets$ and skipping nets already in $visited\_nets$
		\FORALL{first reachable latch $v$ in $F$}
		\STATE Add directed edge $(u, v)$ to $G$
		\IF{color($u$) $=$ color($v$)}
		\STATE \textbf{Error:} Two-color violation at ($u$, $v$)
		\ENDIF
		\ENDFOR
		\ENDFOR
		\RETURN $G$
	\end{algorithmic}
\end{algorithm}

We apply this method through OpenDB, a physical design database used within OpenROAD~\cite{OpenDB_GitHub} and OpenROAD's API~\cite{OpenROADAPI}, where we construct $G = (V, E)$ directly from the placed netlist, as shown in Algorithm~\ref{alg:buildgraph}. \textsc{latchgraph} first populates $V$ by iterating over all instances in the design database and identifying sequential elements. For each sequential element, a DFS is performed backwards along the clock net to the input port to determine whether the element is clocked by $\Phi_1$ or $\Phi_2$. For clock-gated designs, the DFS for clock domain accounts for the \textit{AND} gate by tracing through the gate's clock input, ensuring the original clock phase is recovered. A second DFS is then performed forward through combinational logic from each latch stopping at the first reachable sequential element, and each such reachable latch $v$ is added as a directed edge $(u, v)$ to $G$, producing a structural latch-to-latch connectivity graph suitable for two-coloring verification. Within each DFS traversal from latch $u$, the $visited\_nets$ set prevents re-traversal of the same net, ensuring termination in the presence of combinational loops and FSM feedback paths while still allowing $G$ to represent cycles. The two-coloring check is applied after adding the directed edge $(u, v)$ to $G$; it verifies the graph by checking that for every directed edge $(u, v)$, the clock domains of $u$ and $v$ differ, a single condition that verifies all three of Wolf's two-color constraints~\cite{Wolf2009VLSI}. Constraint 1 is directly enforced: every latch boundary transfers data between opposite phase domains.  Constraint 2 is implicitly enforced by the stopping condition: because the forward DFS halts at the \textit{first} reachable latch, there is no intermediate latch between $u$ and $v$, meaning the data signal arriving at $v$'s input originates directly from $u$'s phase domain. Constraint 3 is implicitly checked by \textsc{latchgraph} construction since edges are built by DFS through combinational logic before stopping at the next latch, preventing cross-phase signal mixing within any logic cone.


\section{Experimental Setup}
\label{sec:expsetup}

For our experimental setup, we use modified OpenROAD (commit 22a4b4cbaa) and the two-phase designs are evaluated against the baseline using the default ORFS workflow. Two-phase SDC constraints are set with $180^\circ$ offset waveforms and a \textit{0.49} duty cycle. We use the default \textit{config.mk} settings for each design and the Google Skywater high density library ~\cite{SkyWater130PDK}. Results are analyzed at the nominal TT 25C 1.8V corner. Reported metrics include total power (with a 10\% switching activity rate on signal inputs, a common default for early-stage power estimation), clock period, max time borrow, actual time borrow, gated clock power for clock-gated designs, clock skew, sequential cell count, combinational cell count, total cell count, design area, and wire length. The flip-flop baseline designs are synthesized using the default ORFS ABC speed script, which applies standard optimization passes without minimum-delay or minimum-area retiming and instead focuses on optimizing delay; the two-phase script described below incorporates iterative retiming passes tailored for latch pair redistribution.

\textbf{ABC Synthesis Script.}
Our ABC script employs an iterative retime optimization strategy targeting logic depth reduction and timing closure. The network is first converted into an And-Inverter Graph (AIG) via structural hashing (\textit{strash}) and \textit{balance}, with an initial register redistribution performed by \textit{dretime}.

Each of four optimization iterations follows the same structure: the network is re-strashed and depth-balanced, functionally equivalent nodes are merged via \textit{dfraig}, structural optimization is applied with \textit{dc2~-b~-p}, and deep rewriting and refactoring passes (\textit{drw~-z}, \textit{drf~-z}) further reduce area under a zero-cost acceptance policy. Retiming with minimum-area and minimum-delay (\textit{retime~-M~5}) is then applied to the optimized network, where a depth-reduced AIG maximizes the freedom of register movement across logic paths.

A final cleanup pass applies another round of rewriting and balancing without retiming, ensuring the network presented to the mapper is depth-minimal. Redundant sequential elements are removed with \textit{scleanup} before technology mapping (\textit{map}). Timing is finalized through topological ordering (\textit{topo}), static timing analysis (\textit{stime~-c}), and buffer/gate sizing (\textit{buffer~-c}, \textit{upsize~-c}, \textit{dnsize~-c}).



\section{Experimental Results}
\label{expresults}
\begin{table*}[htbp]
	\centering
	\caption[Experimental Results]{Experimental Results\\[2pt]
	{\small All timing values are in ns. All power values are reported with 10\% switching activity rate.}}
	\label{tab:results}
	\renewcommand{\arraystretch}{1.15}
	\resizebox{\textwidth}{!}{%
		\begin{tabular}{|l|l|c|c|c|c|c|c|c|c|c|c|c|c|c|}
			\hline
			\textbf{Design}            &
			\textbf{Variant}           &
			\textbf{Per.}              &
			\textbf{Max TB$^\dagger$}  &
			\textbf{Act. TB$^\dagger$} &
			\textbf{Set. Slk}          &
			\textbf{Hld. Slk}          &
			\textbf{Skew}              &
			\textbf{G. Pwr ($\mu$W)}   &
			\textbf{Tot. Pwr (mW)}     &
			\textbf{Seq.}              &
			\textbf{Comb.}             &
			\textbf{Tot. Cells}        &
			\textbf{Area ($\mu$m$^2$)} &
			\textbf{Wire (mm)}                                                                                                                               \\
			\hline\hline
			\multirow{3}{*}{\textbf{RISCV32I}}
			                           & Clock-Gated & 6.4  & 3.08 & 2.48 & ---    & 0.023 & 2.63 & 1.658 & 29.1 & 1133  & 7613  & 31165  & 78664  & 597.73  \\ \cline{2-15}
			                           & Recirc Mux  & 6.4  & 3.05 & 2.88 & ---    & 0.286 & 0.18 & ---   & 51.7 & 4214  & 8184  & 52904  & 137259 & 1068.48 \\ \cline{2-15}
			                           & Baseline    & 6.4  & ---  & ---  & -0.515 & 0.612 & 0.07 & ---   & 21.3 & 1056  & 3502  & 18735  & 70025  & 530.45  \\
			\hline\hline
			\multirow{3}{*}{\textbf{GCD}}
			                           & Clock-Gated & 3.99 & 1.76 & 1.1  & ---    & 0.049 & 1.16 & 158.0 & 2.30 & 75    & 268   & 1382   & 4155   & 14.58   \\ \cline{2-15}
			                           & Recirc Mux  & 3.99 & ---  & ---  & 0.101  & 0.310 & 0.31 & ---   & 4.53 & 164   & 321   & 2141   & 5779   & 23.68   \\ \cline{2-15}
			                           & Baseline    & 3.99 & ---  & ---  & 0.307  & 0.333 & 0.0  & ---   & 1.36 & 35    & 187   & 892    & 3040   & 10.06   \\
			\hline\hline
			\multirow{3}{*}{\textbf{AES}}
			                           & Clock-Gated & 4.7  & 2.06 & 1.78 & ---    & 0.141 & 1.6  & 457.0 & 2700 & 1237  & 10879 & 73747  & 118454 & 1485.00 \\ \cline{2-15}
			                           & Recirc Mux  & 4.7  & 2.2  & 1.83 & ---    & 0.225 & 0.28 & ---   & 2560 & 1599  & 10711 & 77594  & 120299 & 1509.32 \\ \cline{2-15}
			                           & Baseline    & 4.7  & ---  & ---  & 0.076  & 0.045 & 0.21 & ---   & 420  & 582   & 10722 & 68410  & 115208 & 1494.14 \\
			\hline\hline
			\multirow{3}{*}{\textbf{JPEG}}
			                           & Clock-Gated & 5.5  & 2.64 & 1.46 & ---    & 0.060 & 0.28 & 11200 & 1750 & 9666  & 60178 & 211315 & 588481 & 3463.99 \\ \cline{2-15}
			                           & Recirc Mux* & 5.5  & ---  & ---  & ---    & ---   & ---  & ---   & ---  & 17325 & 89887 & ---    & ---    & ---     \\ \cline{2-15}
			                           & Baseline    & 5.5  & ---  & ---  & 0.037  & 0.345 & 0.1  & ---   & 485  & 4390  & 27031 & 118475 & 452808 & 2489.21 \\
			\hline
			\multicolumn{15}{l}{%
				\footnotesize $^\dagger$Abbreviated as \textbf{T}ime \textbf{B}orrow. \quad
				\footnotesize $^{*}$JPEG recirculation mux variant not reported due to failures in routing from congestion.
			}                                                                                                                                                \\
			\multicolumn{15}{l}{%
				\footnotesize Per.\ = Period, Act.\ = Actual, Set.\ Slk = Setup Slack, Hld.\ Slk = Hold Slack, G.\ = Gated, Tot.\ = Total, Seq.\ = Sequential Cells, Comb.\ = Combinational Cells.
			}                                                                                                                                                \\
		\end{tabular}%
	}
\end{table*}

\begin{figure*}[htbp]
	\centering
	\begin{minipage}[b]{0.45\textwidth}
		\centering
		\includegraphics[width=\textwidth]{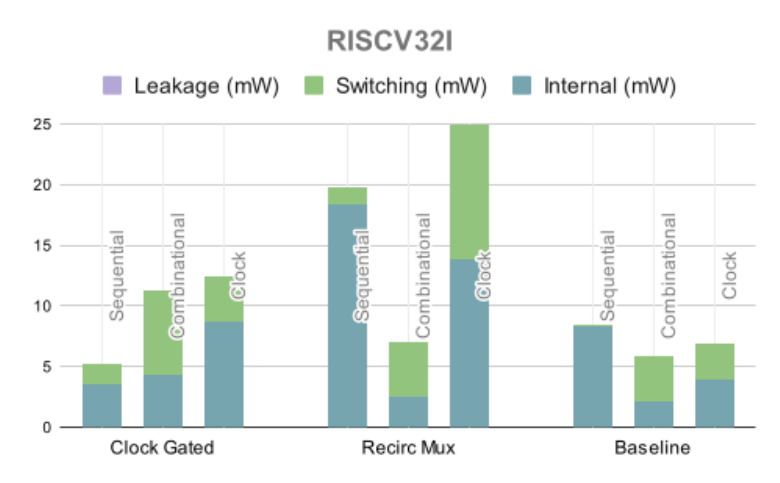}
	\end{minipage}
	\hfill
	\begin{minipage}[b]{0.45\textwidth}
		\centering
		\includegraphics[width=\textwidth]{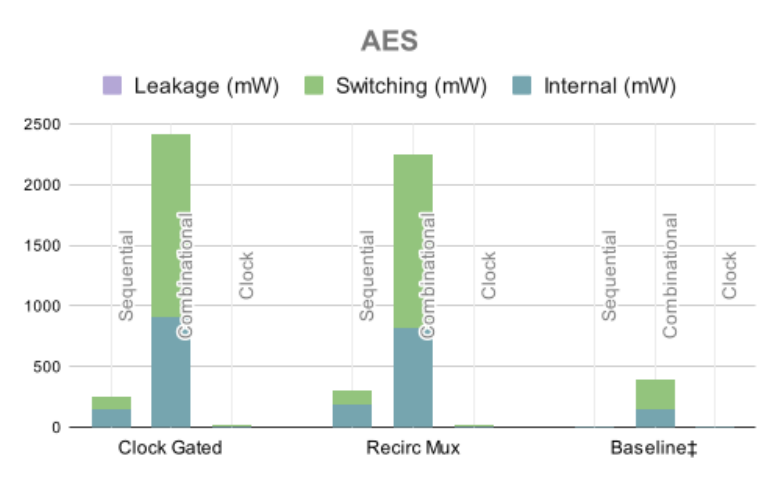}
	\end{minipage}
	\\
	\hfill{\footnotesize $^{\ddagger}$Sequential and clock power reported 11.2\,mW and 9.07\,mW respectively.}
	\caption{Power trade-off of two-phase latch variants over flip-flops; clock-gating reduces total power compared to recirculation muxes on RISCV32I, but provides no gain to AES where combinational switching dominates.}
	\label{fig:power-graphs}
\end{figure*}

Table~\ref{tab:results} compares the two-phase designs against flip-flop baselines for RISCV32I, GCD, AES, and JPEG at identical target periods after running end-to-end RTL-to-GDS flows. All designs passed two-coloring, making each design a valid two-phase system. Two-phase clocking enables timing closure through time borrowing at the cost of increased power and latch count. Most notably, the RISCV32I baseline fails timing at 6.4\,ns with a setup slack of $-0.515$\,ns, while both two-phase variants close timing at the same period by borrowing up to 2.88\,ns. For designs that already meet timing in baseline (GCD, AES), two-phase clocking provides additional timing margin. JPEG's clock-gated variant demonstrates conservative borrowing, using only 1.46\,ns of the available 2.64\,ns, suggesting JPEG has potential to increase performance by pushing time borrowing further.

\subsection{Power and Area Analysis}
Regarding power and area, the cost of timing closure is significant, as shown in Figure~\ref{fig:power-graphs}. RISCV32I total power increased $+36.6\%$ from baseline (21.3\,mW) to clock-gated (29.1\,mW) and $+142.7\%$ to recirculation mux (51.7\,mW). Clock-gated adds $+7.3\%$ sequential cells over baseline (1133 vs.\ 1056), while recirculation mux adds $+299.1\%$ (4214 vs.\ 1056). JPEG shows a similar pattern: clock-gated succeeds with 9666 cells and $+260.8\%$ power over baseline (1750\,mW vs.\ 485\,mW), while recirculation mux failed due to routing congestion from its 17325 latch count ($+294.6\%$ over baseline). AES is the outlier: clock-gated and recirculation mux variants are substantially worse in power than baseline ($+542.9\%$ and $+509.5\%$), yet both variants differ from each other by only $+5.5\%$ despite a $-22.6\%$ difference in latch count (1237 vs.\ 1599), suggesting power is dominated by combinational logic rather than latch overhead. Table~\ref{tab:results} shows the power increase between baseline and two-phase designs is largely driven by three factors: higher total cell counts from latch duplication, larger design areas from the additional sequential and clock-gating logic, and longer wire lengths from the dual clock trees and increased fanout. As observed for RISCV32I, combinational cells grow from 3502 to 7613 (clock-gated) and 8184 (recirculation mux), with wire length increasing from 530.45\,mm to 597.73\,mm and 1068.48\,mm respectively, directly correlating with the observed power increase. Notably, RISCV32I clock-gated variant achieves the lowest power increase among two-phase designs of $+36.6\%$ and the smallest latch count increase $+7.3\%$, suggesting that minimum-area retiming produced effective results. The elevated clock skew in the RISCV32I clock-gated variant (2.63\,ns vs.\ 0.18\,ns for recirculation mux) is attributable to the \textit{AND} gate inserted in the $\Phi_2$ clock path for latch enable gating, which introduces an asymmetric insertion delay relative to the ungated $\Phi_1$ clock tree; this asymmetry is specific to the clock-gated topology and is absent in the recirculation mux variant, where both clock paths remain structurally symmetric. AES, however, does not follow this trend: all three variants share nearly identical design areas ($\sim$115K--120K\,$\mu$m$^2$) and wire lengths ($\sim$1485--1509\,mm), suggesting that the dominant power contribution in AES comes from its deeply combinational datapath rather than sequential cell or interconnect overhead, as dual-clock operation amplifies switching activity through the unrolled cipher rounds, making cell count and wire length poor predictors of power for this design.

\textbf{Clock-Gated vs Recirculation Muxes} Among two-phase variants, clock-gating yields fewer latches and lower power than recirculation mux. For RISCV32I, clock-gating reduces total power by $-43.7\%$ (29.1\,mW vs.\ 51.7\,mW) and latch count by $-73.1\%$ (1133 vs.\ 4214), with GCD showing a similar trend ($-49.2\%$ power, $-54.3\%$ latches). AES is the exception, where variants differ by only $+5.5\%$ in power despite a $-22.6\%$ difference in latch count.


\section{Conclusion}
This paper presents the first fully automated two-phase clocking flow in ORFS, converting flip-flop RTL to latch-based designs via Yosys mapping, ABC retiming, dual clock tree synthesis, and two-coloring validation. While two-phase clocking improves timing closure through time borrowing, it incurs significant power overhead over baseline across all designs. The strongest result is RISCV32I clock-gated, which recovers timing closure at 6.4\,ns from a failing baseline (setup slack of $-0.515$\,ns), while incurring only $+36.6\%$ power overhead, the smallest increase of any two-phase design across all benchmarks. Among the two variants, clock-gating reduces power by $-29.2\%$ and latch count by $-50.0\%$ on average over recirculation mux, making it the more practical choice when power and area are constrained. These results demonstrate that two-phase clocking is now viable in an open-source RTL-to-GDS flow.




\bibliographystyle{ACM-Reference-Format}
\bibliography{software,twophase-base}

\end{document}